\documentclass{amsart}
\newtheorem{thm}{Theorem}[section]
\newtheorem{cor}[thm]{Corollary}

\theoremstyle{definition}
\newtheorem{defn}[thm]{Definition}
\theoremstyle{remark}

\numberwithin{equation}{section}
\begin{document}

\title{Speed Limits in General Relativity}
\author{Robert J. Low}
\address{Mathematics Division, School of Mathematical and Information Sciences,
Coventry University, Priory Street, Coventry CV1 5FB, U.K.}
\email{r.low@coventry.ac.uk}

\subjclass{Primary 83C05; Secondary 83C35, 83C55}
\keywords{Cauchy problem, warp drive}
\date{}

%%% ----------------------------------------------------------------------

\begin{abstract}
Some standard results on the initial value problem of general
relativity in matter are reviewed. These results are applied first
to show that in a well defined sense, finite perturbations in the
gravitational field travel no faster than light, and second to show
that it is impossible to construct a warp drive as considered by
Alcubierre (1994) in the absence of exotic matter.
\end{abstract}

%%% ----------------------------------------------------------------------
\maketitle
%%% ----------------------------------------------------------------------
\section{Introduction}
This paper is divided into four sections. In the first, I
review some standard results on the initial value problem of
general relativity, both for vacuum space-times and for
space-times containing matter. In the second section, I
consider the propagation of gravitational disturbances. It is
well known that in the limit of small perturbations, such
disturbances travel along null geodesics, and so
gravitational waves travel at light speed. An initial value
approach to non-infinitesimal perturbations will be developed
that shows that, in a well-defined sense, finite changes in
the gravitational field propagate no faster than light speed.
In the third section, I consider the problem of whether it is
possible to construct a `warp drive', such as is described in
Alcubierre (1994). This necessitates some way of describing
what it means to do such a thing in a deterministic theory;
after developing such a notion I show that it is impossible
to build a warp drive if all matter present satisfies the
dominant energy condition. Alcubierre's conjecture that
exotic matter is required for the construction of a warp
drive is therefore seen to be correct. Finally, I give a
brief discussion of the results.

\section{The Initial Value Problem}
In this section I will describe some standard results.
Outlines of the proofs may be found in Chapter 10 of Wald
(1984); detailed proofs are in Chapter 7 of Hawking and
Ellis (1973), and a more geometric approach which sets up a
Hamiltonian formulation for the vacuum case is presented in
Fischer and Marsden (1976).

Let $(M,g_{\alpha \beta})$ be a Lorentz manifold satisfying the
Einstein equations, so that $G_{\alpha \beta}=0$. Then if $\Sigma$ is
a spacelike surface in $M$, the metric $g_{\alpha \beta}$ induces on
$\Sigma$ a Riemannian three-metric, $h_{ab}$, and a
symmetric tensor $K_{ab}$, the second fundamental form of
$\Sigma$. The following constraints are implied by the
Gauss-Codazzi relations:
\begin{equation} \label{vaccon}
\begin{array}{rcl}
D_bK^b_{\;\;a} - D_aK^b_{\;\;b} &=& 0 \\
^{(3)}R +(K^a_{\;\;a})^2 - K_{ab}K^{ab} &=& 0
\end{array}
\end{equation}
where $D_a$ is the covariant derivative on $\Sigma$ defined by
$h_{ab}$, and $^{(3)}R$ is the scalar curvature of $h$.

Define a vacuum initial data set to be a triple
$(\Sigma,h_{ab},K_{ab})$ where $\Sigma$ is a smooth three
manifold, $h_{ab}$ is a Riemannian metric on $\Sigma$, and
$K_{ab}$ is a symmetric tensor on $\Sigma$ satisfying the
constraint equations (\ref{vaccon}).
Then we have the following theorem:
\begin{thm} \label{max}
Let $(\Sigma,h_{ab},K_{ab})$ be a vacuum initial data set.
Then
\begin{enumerate}
\item there exists a unique smooth space-time $(M,g_{\alpha \beta})$
known as the maximal development of $(\Sigma, h_{ab},K_{ab})$
and a smooth embedding $i:\Sigma \rightarrow M$ such that
$M$ is globally hyperbolic with Cauchy surface $i(M)$, the
metric and second fundamental form induced on $i(\Sigma)$ by
$g_{\alpha \beta}$ are, respectively, $i_*(g_{ab})$ and $i_*(K_{ab})$.
Furthermore, any other space-time satisfying these conditions
may be mapped isometrically into $(M,g_{\alpha \beta})$.
\item if $(\Sigma',h'_{ab},K'_{ab})$ is a vacuum initial data set
with maximal development $(M',g'_{\alpha \beta})$ and embedding
$i':\Sigma' \rightarrow M'$, and there is a
diffeomorphism from $S \subseteq \Sigma$ to
$S' \subseteq \Sigma'$ taking $(h_{ab},K_{ab})$ on $S$ to
$(h'_{ab},K'_{ab})$ on $S'$, then the Cauchy development
$D(i(S))$ in $M$ is isometric to the Cauchy development
$D(i'(S'))$ in $M'$.
\item the metric $g_{\alpha \beta}$ on $M$ depends continuously on the
vacuum initial data $(h_{ab},K_{ab})$ on $\Sigma$.
\qed
\end{enumerate}
\end{thm}
It is worth noting that $M$ need not be an inextensible
space-time; however, if $j:M \rightarrow M'$ is an isometry
onto its image and fails to be surjective, $j(i(\Sigma))$
cannot be a Cauchy surface for $M'$. Fischer (1995) has an
interesting discussion of this and related issues.

Of more relevance here, however, is the following simple
consequence, obtained by letting $S=\Sigma$ in part (ii) of
the above theorem.
\begin{cor} \label{dep}
Let $(\Sigma,h_{ab},K_{ab})$ and $(\Sigma',h'_{ab},K'_{ab})$
be vacuum initial data sets such with maximal developments
$(M,g_{\alpha \beta})$ and $M',g'_{\alpha \beta})$ respectively, and suppose
that there is a diffeomorphism $j$ mapping $\Sigma$ to a subset of
$\Sigma'$ which takes $(h_{ab},K_{ab})$ to
$(h'_{ab},K'_{ab})$. Then $M$ is isometric to $D(i'(j(\Sigma)))$
in $M'$.
\qed
\end{cor}
Similar results hold for the Einstein equations with certain
forms of matter present.

In particular, suppose we have matter described by a vector
$\boldsymbol{u}$ satisfying a quasi-linear symmetric hyperbolic
system, i.e. a differential equation of the form
\begin{equation} \label{evol}
A^{\alpha}\nabla_{\alpha}\boldsymbol{u} +B\boldsymbol{u} =
\boldsymbol{c}
\end{equation}
where
\begin{enumerate}
\item $A^{\alpha}$ is a four-vector of symmetric matrices,
\item $B$ is a matrix, and
\item $\boldsymbol{c}$ is a vector
\end{enumerate}
where $A^{\alpha}$, $B$ and $\boldsymbol{c}$ can all depend on
$\boldsymbol{u}$ as well as position. We will mostly be interested
in the case where the ray cone for the system (the boundary of 
the set of vectors  $v^\alpha$ making $v_\alpha A^\alpha$ positive 
definite) lies inside the light cone.

It is a standard result (Courant and Hilbert, 1962) that this
includes the case of matter describe by a second order
hyperbolic system where the bicharacteristics are timelike or
null. The physical significance of this constraint on
$A^{\alpha}$ is that the bicharacteristics of the matter
equations are causal curves, so that the matter cannot
support the transfer of information at speed exceeding that
of light.

Suppose also that the stress tensor of the matter present, $T_{\alpha
\beta}$, depends only on $\boldsymbol{u}$ and $g^{\alpha \beta}$.

We will be concerned with the issue of whether it is possible to
construct a warp drive in the absence of exotic matter; to this
end, we require a definition of exotic matter;

\begin{defn} \label{exotic}
A matter field will be called \textit{physically reasonable} if it
satisfies a symmetric hyperbolic evolution equation and its
stress-energy tensor satisfies the dominant energy condition;
otherwise it will be called \textit{exotic}.
\end{defn}

It now follows that if $\Sigma$ is a spacelike surface in
$(M,g_{\alpha \beta})$ where $G_{\alpha \beta}= 8\pi
T_{\alpha \beta}$, then again denoting the induced Riemannian
metric on $\Sigma$ by $h_{ab}$ and its second fundamental
form by $K_{ab}$, we find
\begin{equation} \label{matcon}
\begin{array}{rcl}
D_bK^b_{\;\;a} - D_aK^b_{\;\;b} &=& -8\pi J_b\\
^{(3)}R +(K^a_{\;\;a})^2 - K_{ab}K^{ab} &=& 16\pi\rho
\end{array}
\end{equation}
where $\rho = T_{\alpha \beta}n^\alpha n^\beta$ and $J_a$ is
the projection to $\Sigma$ of $T_{\alpha \beta}n^\beta$, $n^\alpha$
being the future pointing normal to $\Sigma$.

We can now define an initial data set
$(\Sigma,h_{ab},K_{ab},J_a,\rho)$ as a three manifold with
a three metric $h_{ab}$, a symmetric tensor $K_{ab}$, a
vector $J_a$ and a scalar $\rho$  satisfying the above
equations. There follows a theorem analogous to Theorem
\ref{max} above, except that now the metric $g_{\alpha
\beta}$ satisfies $G_{\alpha \beta}=8\pi T_{\alpha \beta}$
where $T_{\alpha \beta}$ is the stress tensor of the matter
field described by $u$ and satisfying the evolution equation
(\ref{evol}). Similarly, there is an analogous corollary to
Corollary \ref{dep}.

Explicitly, we have
\begin{thm} \label{maxmat}
Let $(\Sigma,h_{ab},K_{ab},J_a,\rho)$ be an initial data set. Then
\begin{enumerate}
\item there exists a unique smooth space-time $(M,g_{\alpha
\beta})$ known as the maximal development of $(\Sigma,
h_{ab},K_{ab},J_a,\rho)$ and a smooth embedding $i:\Sigma
\rightarrow M$ such that $M$ is globally hyperbolic with
Cauchy surface $i(M)$, the metric and second fundamental
form induced on $i(\Sigma)$ by $g_{\alpha \beta}$ are,
respectively, $i_*(g_{ab})$ and $i_*(K_{ab})$, and $J_a$
and $\rho$ are induced on $\Sigma$ by a matter field
satisfying (\ref{evol}). Furthermore, any other space-time
satisfying these conditions may be mapped isometrically in
$(M,g_{\alpha \beta})$.
\item if $(\Sigma',h'_{ab},K'_{ab},J'_a,\rho')$ is a vacuum
initial data set with maximal development $(M',g'_{\alpha \beta})$
and embedding $i':\Sigma' \rightarrow M'$, and there is a
diffeomorphism from $S \subseteq \Sigma$ to  $S' \subseteq
\Sigma'$ taking $(h_{ab},K_{ab},J_a,\rho)$ on $S$ to
$(h'_{ab},K'_{ab},J'_a,\rho')$ on $S'$, then the Cauchy
development $D(i(S))$ in $M$ is isometric to the Cauchy
development $D(i'(S'))$ in $M'$.
\item the metric $g_{\alpha \beta}$ on $M$ depends
continuously on the initial data $(h_{ab},K_{ab},J_a,\rho)$ on
$\Sigma$. \qed
\end{enumerate}
\end{thm}
\noindent
with corollary
\begin{cor} \label{matdep}
Let $(\Sigma,h_{ab},K_{ab},J_a,\rho)$ and
$(\Sigma',h'_{ab},K'_{ab},J'_a,\rho')$ be initial data sets
such with maximal developments $(M,g_{\alpha \beta})$ and
$(M',g'_{\alpha \beta})$ respectively, and suppose that there is a
diffeomorphism $j$ mapping $\Sigma$ to a subset of
$\Sigma'$ which takes $(h_{ab},K_{ab},J_a,\rho)$ to
$(h'_{ab},K'_{ab},J'_a,\rho')$. Then $M$ is isometric to
$D(i'(j(\Sigma)))$ in $M'$.
\qed
\end{cor}
It is perhaps worth noting here that this theorem and corollary
are true for a \textit{fixed} equation of the form (\ref{evol}). One 
could have identical initial conditions with non-isometric
maximal developments if the physical properties of the matter
fields involved, and hence the evolution equations, were different.
In the sequel, we will consider only the effects of changing the
initial conditions, while keeping the evolution equation fixed.

\section{Evolution of gravitational perturbations}
It is well known that linear perturbations of the metric on an
Einstein vacuum satisfy a wave equation, and that the
bicharacteristics of this wave equation are the null geodesics of
the background space time (Wald 1984). In this sense, then, small
perturbations in the gravitational field propagate at the speed
of light.

However, it is somewhat more problematic to describe what happens
with a finite perturbation of the metric. The difficulty is
two-fold. First, there is the problem of deciding just what the
perturbation is---the Einstein equations are non-linear, so there
is no natural splitting into background field and disturbance.
Second, since the metric itself determines the null geodesics,
what can it mean to say that perturbations in the metric travel
at light speed?

Let us approach this by first recasting the case of
infinitesimal perturbations. So, let $\Sigma$ be a Cauchy
surface for $(M,g_{\alpha \beta})$, and let $K$ be a
submanifold of $\Sigma$. If we make an infinitesimal
change to the vacuum initial data on $K$, while leaving
that on $\Sigma \setminus K$ unchanged, then the metric is
unchanged on $D(\Sigma \setminus K)$, since the
perturbation travels at light speed and therefore cannot
reach any point outside $I(K)$.

So now consider the effect of changing the initial data on
$K$ by a finite amount. We can no longer split the
space-time into a background metric with a perturbation, but
must proceed as follows: denote the first vacuum initial
data set by
$(\Sigma,h_{ab},K_{ab})$, and the altered one by
$(\Sigma,h'_{ab},K'_{ab})$, noting that $(h_{ab},K_{ab})$
and $(h'_{ab},K'_{ab})$ agree outside $K$. Also, denote the
first maximal development by $(M,g_{\alpha \beta})$ and the
other by $(M',g'_{\alpha \beta})$.

By restricting we obtain an initial data set
$(\Sigma \setminus K,h_{ab},K_{ab})$, with associated
maximal development $(M^0,g^0_{\alpha \beta})$. It follows
immediately from Corollary \ref{dep} that there is an
isometry from $M^0$ to $D(i(\Sigma \setminus K))$ in $M$, and
also to $D(i'(\Sigma \setminus K))$ in $M'$. We thus obtain
\begin{cor} \label{vacisom}
With all quantities as defined above, there is an isometry
from $D(i(\Sigma \setminus K)) \subset M$ to $D(i'(\Sigma
\setminus K)) \subset M'$.
\qed
\end{cor}
Since in $M$ no influence can propagate from a point of $K$
to a point of $D(i(\Sigma \setminus K))$ along a causal
curve, and $M'$ differs from $M$ only outside
$D(i(\Sigma \setminus K)) \equiv D(i'(\Sigma \setminus K))$,
we can interpret this result as stating that no
gravitational influence propagates faster than light.

\section{Warp Drive}
Let us now proceed to the somewhat more subtle question of
whether it is possible to construct a warp drive. Analysis
(Pfenning and Ford 1997) has shown that the
particular metric considered by Alcubierre (1994) is
unphysical; there remains the question of whether there is a
physically reasonable metric exhibiting analogous properties.

Before attempting to answer the question, it is necessary to
find a way of asking it. The difficulty is that in general
relativity the metric of space-time is already fixed: there
is no obvious sense in which one can make a decision which
would change the metric to one's future from what it would
otherwise have been.

To motivate the sequel, consider the case of a test field evolving
on a fixed space-time background, and let $S$ be a Cauchy
surface in the fixed space-time. Then if the initial data
determining the field on $S$ differ only in $I^+(p)
\cap S$, for some event $p$, we might reasonably regard
this difference as due to a decision made at $p$. The question
is, how do we alter this to cope with the case where the initial
data determines the space-time?

Consider the space-time $(M,g_{\alpha \beta})$, defined as the
maximal development of some initial data set
$(\Sigma,h_{ab},K_{ab},J_a,\rho)$, where the matter fields satisfy
some equation of the form (\ref{evol}) with the ray cones lying
inside the light cones. Let $p$ be some point to the past of
$i(\Sigma)$, and let $K = J^+(p) \cap \Sigma$. Now, consider an
initial data set that agrees with this one on $\Sigma \setminus K$,
but differs inside $K$. This will have a maximal development
$(M',g'_{\alpha \beta})$. As in the vacuum case, we obtain from
Corollary \ref{matdep} the further result
\begin{cor} \label{matisom}
With all quantities as defined above, there is an isometry
from $D(i(\Sigma \setminus K)) \subset M$ to $D(i'(\Sigma
\setminus K)) \subset M'$.
\qed
\end{cor}
All the following results will hold if we model the decision to
atttempt to construct a warp drive by such an alteration in the
initial data on some Cauchy surface to the future of the decision.

If we wish, we can consider only changes in the initial data
which result in a (near) isometry from $I^-(i(\Sigma \setminus K))$
in $M$ to $I^-(i'(\Sigma \setminus K))$ in $M'$, so that the space-times
initially look as nearly isometric as we wish, and then begin to
diverge noticeably to the future of $p$. However, since the
result to be established will hold for unrestricted changes to the
initial data in $K$, it will \textit{a forteriori} hold under any
constraint made to those changes. We will therefore eschew any
further discussion of such constraints.

Note that no requirement has been made on the relationship
between $I^+(i(\Sigma))$ in $M$ and $I^+(i'(\Sigma))$ in $M'$; indeed, 
we should wish it to be possible for the
mapping to be far from an isometry there. The kind of
situation this should model is where some large-scale
machinery has been constructed, and the decision of
whether or not to set it in motion is made at some event
$p$.

So now, let us consider the consequences of this for the
attempt to construct a warp drive. First, consider the
space-time $M$. Let $A$ be a star (modelled as a point)
whose world-line, $L$, intersects $i(\Sigma \setminus K)$. Then
there are two possibilities: either $L$ remains inside
$D^+(i(\Sigma \setminus K))$, or it eventually leaves this
domain of dependence. In the former case, there is no causal
path in $M$ from $p$ to $L$; in the latter, it is possible
to travel from $p$ to $L$ and the earliest point on $L$ one
can arrive at is the intersection of $L$ with the boundary
of $D^+(i(\Sigma \setminus K))$.

Now, there is an isometry from $D(i(\Sigma \setminus K)$ to
$D(i'(\Sigma \setminus K))$; it follows that also in $M'$
either the worldline of $A$ remains within the domain of
dependence of $i'(\Sigma \setminus K)$, or that the interval
inside this domain of dependence isometric to that inside
$D(i(\Sigma \setminus K))$. Thus in neither case is it
possible to reach $L$ any sooner than it is in $M$, in the
sense that one cannot reach an earlier point on the
worldline of $L$. It may, however, be possible to arrange
space-time in such a way that a slower space-ship can arrive
with less elapsed proper time; more interestingly, it may be
possible to arrange space-time in such a way that the return
trip may be made short as measured by an observer who
remains on the Earth. This possibility is considered in more
detail in Krasnikov (1998).

We see, then, that with this notion of what it means to
`change the metric', and if all matter satisfies an evolution
equation of the form (\ref{evol}) with the ray cones inside
the light cones, one cannot construct a warp drive which has
the effect of making it possible to travel `faster than
light'. However, there remains the case of what happens if
the matter is not of this type: is it possible for physically
reasonable matter to avoid the consequences developed above?

Still assuming that any matter fields are described by an evolution
equation of the form \ref{evol}, we see that in order
for a warp drive to be effective---i.e. for it to be possible
to change the initial data on $K$ in such a way that the
portion of $L$ in $I^+(i'(\Sigma \setminus K))$ is shorter
than that in $I^+(i(\Sigma \setminus K))$, the ray cone of
the evolution equation (\ref{evol}) cannot lie inside the
null cone of the metric $g_{\alpha \beta}$. But this means
that the matter field can support the propagation of a signal
at a speed exceeding that of light; however, (Hawking and
Ellis, 1973), this is only possible if the matter violates
the dominant energy condition. It therefore follows that a
warp drive can only be constructed in the presence of exotic
matter, confirming the suspicion voiced in Alcubierre (1994).

This result can be made rather more vigorous; as long as the
initial data on $\Sigma \setminus K$ is unchanged, and the
matter satisfies the evolution equation (\ref{evol}) with ray
cones lying inside the light cones, $I^+(\Sigma \setminus
K))$ is unchanged. Thus, whatever constraint one chooses on
the data in $K$ to provide some interpretation of a free will
decision, the result will hold.

In fact, we can go still further. In the case where $L$
eventually departs $I^+(i(\Sigma \setminus K))$ in $M$, there
is a compact set $V$ containing $L \cap I^+(i(\Sigma
\setminus K))$. Then as long as the initial data on
$\Sigma \setminus K$ changes sufficiently little, the metric
on $V$ will also be changed by little, by the continuity
theorem given in Hawking and Ellis (1973). So, even if we
allow the initial data outside $K$ to change, by making the
change sufficiently small, we can ensure that the proper time
the star $A$ spends in $I^+(i'(\Sigma \setminus K))$ is very
near that in the original space-time $M$.

So, gathering this together, and recalling that those matter 
fields whose evolution is not determined by a symmetric 
hyperbolic system or which fail to satisfy the dominant energy 
condition are by definition (\ref{exotic}) exotic, we have the result:
\begin{thm}
In the absence of exotic matter, it is impossible to construct a
warp-drive. \qed
\end{thm}

\section{Conclusions and Discussion}
We have seen that in the absence of exotic matter, the speed of
light is in a well-defined sense a speed limit in general
relativity. In order to do this it was necessary to make explicit
the sense in which one might talk about the propagation of
perturbations in space-time, rather than that of test fields or
test particles. This approach could be applied both in the case of
vacuum space-times and those containing matter.

It is also possible to use the ideas exposed above to extend
some other discussions beyond the realm of the behaviour of
test particles.

For example, it is clear that a test particle inside the event
horizon of maximally extended Schwarzschild space, $\mathcal{S}$,
cannot escape. It is less obvious that a non-infinitesimal
gravitational wave or physical particle cannot do so, since such
objects affect the space-time itself. However, if we consider a
Cauchy surface $\Sigma$ for $\mathcal{S}$ which enters region II
(in the notation of Hawking and Ellis, 1973), we can consider an
alternative space-time defined by changing the initial data on
$\Sigma$ on a subset, $K$, whose closure lies inside region II.
From the above theorems, $D^+(\Sigma \setminus K)$ is unchanged by
this alteration, so that some neighbourhood of the event horizon
and the entirety of the external region to the future of $\Sigma$
are unchanged. This gives a simple way of seeing that unless the
collapsing matter is exotic, once it has fallen inside the apparent
horizon, the subsequent external solution is just external Schwarzschild.
Similar comments will, of course, hold for the Kerr and
Reissner-Nordstr\"om solutions.

This can even be interpreted as a kind of mini-censorship
theorem, telling us that any singularity developing from a
physically reasonable matter distribution inside the apparent
horizon of an initial data surface for a  black hole will lie inside the 
event horizon of the future development; \textit{i.e.} it cannot be
observed from the asumptotically flat region.
(Recall that the classical singularity theorems only tell us
that in the presence of a closed trapped surface, a
singularity must develop---they do not tell us where the
singularity will be.)

\noindent
\textbf{Acknowledgement} I thank the anonymous referees for their
suggestions for the improvement of the presentation of this paper.

% ------------------------------------------------------------------------

\end{document}